\documentclass[12pt]{iopart}

\newcommand{\eq}[1]{\begin{equation}#1\end{equation}}
\newcommand{\dd}{\mathrm{d}}
\newcommand{\ee}{\mathrm{e}}
\usepackage{amssymb}
\usepackage{graphics}
\usepackage{graphicx}
\usepackage{psfrag}
\usepackage{colordvi}

\begin{document}

\title{Entanglement in spin chains with gradients}
\author{Viktor Eisler$^1$, Ferenc Igl\'oi$^{2,3}$ and Ingo Peschel$^1$}
\address{$^1$ Fachbereich Physik, Freie Universit\"at Berlin, Arnimallee
 14, D-14195 Berlin, Germany\\
$^2$ Research Institute for Solid State Physics and Optics, P.O.Box 49,
H-1525 Budapest, Hungary\\
$^3$ Institute of Theoretical Physics, Szeged University, H-6720 Szeged,
Hungary}


\begin{abstract}
We study solvable spin chains where either fields or couplings vary
linearly in space and create a sandwich-like structure of the ground state.
We find that the entanglement entropy between two halves of a chain varies
logarithmically with the interface width. After quenching to a homogeneous
critical system, the entropy grows logarithmically in time in the XX model,
but quadratically in the transverse Ising chain. We explain this behaviour 
and indicate generalizations to other power laws.

\end{abstract}

\section{Introduction}

The entanglement between two parts of a quantum chain has
been the topic of numerous recent studies \cite{Amicoetal08}. For homogeneous
systems, the entanglement entropy $S$ in the ground state
has been found to be a quantity of order one if the system is 
non-critical, while it varies as $\ln L$ if it is critical. Here $L$
is the length of the subsystem, which can be either an interval in an
infinite chain or one-half of a finite chain. The constant
in front of the logarithm is proportional to the number of contact
points between the subsystems and to the central charge of the 
corresponding conformal field theory. In the vicinity of a phase 
transition, $S$ varies as $\ln\xi$, where $\xi$ is the correlation length,
if $L \gg \xi$ \cite{CC04}. There have also been studies of non-homogeneous 
systems. For example, if a defect separates the two subsystems,
the prefactor of $\ln L$ varies with the defect strength in simple 
hopping models \cite{Peschel05,Levine08}, while for interacting electrons it scales 
either to zero or to the non-interacting value for large $L$ \cite{Levine04,
ZhaoPeWa06}. 
The central charge is also modified in chains with random couplings, but
in this case by a constant factor of $\ln2$ \cite{RefaelMoore04,Laflo05,IgloiLin08}.
\par
In the present study we consider non-homogeneous systems of a different
nature. We assume that one parameter in the Hamiltonian varies linearly 
along the chain and consider two cases. In an XX model, we vary the strength 
of a magnetic field in the $z$-direction, while in a transverse Ising (TI) model  
we vary the couplings around the critical value. The first model constitutes
a well-known problem, since it corresponds to free electrons hopping on a chain 
under the influence of a constant electric field. For a sufficiently large system,
the central single-particle levels are then equidistant and form the famous 
Wannier-Stark ladder \cite{Wannier60,Smith71,Saitoh73,CaseLau73,SteyGusman73}. 
The situation has been realized experimentally in optical lattices subject 
to a constant acceleration \cite{Wilkinson96,Niu96}. 
The second model was investigated recently with regard to its critical properties 
\cite{Platini07}. Physically, they have in common that the gradient terms introduce an 
interface into the ground state, and a length scale $\lambda$ which measures its width. 
In the hopping model, this interface separates regions where the system is 
completely full and completely empty, respectively. In the TI model, it separates 
ordered and non-ordered regions. This interface should have a strong influence on 
the entanglement of the regions to the left and right of it. This is, indeed, what 
one finds. The entanglement entropy becomes constant for large $L$ and the asymptotic 
value is determined by $\ln(\lambda)$. Moreover, the deviation from the value in the 
homogeneous system has a scaling form in the variable $(L/\lambda)$.
\par
We also discuss the time evolution if the gradient is suddenly removed. For
the hopping model, it turns out that $S(\lambda,t)$ depends only on the variable 
$(t^2+\lambda^2)$ and thus can be obtained from the ground state entanglement with
the field. This means in particular that it varies logarithmically in time. For the TI 
model, the situation is different and more interesting. Here one finds a
\emph{quadratic} increase of the entanglement with time. Such a behaviour has not
been encountered before in such quenchs, but we show that it can be understood in a 
simple way using the quasiparticle picture of Calabrese and Cardy \cite{CC07}.
\par
In the following section 2 we review briefly the features of the Wannier-Stark
problem. In section 3 we determine the correlation matrix from which S is calculated
and present results for the XX chain in its ground state. In section 4 the TI model 
is considered, again in its ground state, while in section 5 results
for the time evolution after the removal of the gradient are given for both models.
In section 6 we sum up our findings and in an Appendix we comment on more
general spatial inhomogeneities and the derivation of the length scales.
\par
%
  

\section{Wannier-Stark problem}

The problem of lattice electrons in a homogeneous electric field has been the 
subject of many investigations. In the form of a simple one-dimensional tight-binding
model it was studied, for example, in \cite{Feuer52,Saitoh73,CaseLau73,SteyGusman73}.
The equivalent spin one-half XX chain was treated in \cite{Smith71}. The
corresponding eigenvalue equation appears also in the treatment of certain 
reaction-diffusion models \cite{PRS94}. The Hamiltonian is, for a finite
system of $2L$ sites with open ends, 
\eq{
H= -\frac 1 2 \sum_{n=-L+1}^{L-1} (c^{\dag}_n c_{n+1} + c^{\dag}_{n+1} c_{n})
+ h \sum_{n=-L+1}^{L} (n-1/2) c^{\dag}_n c_{n} 
\label{eq:hamXX},}
Here the linear field is chosen such that it goes through zero between
sites 0 and 1 and thus is odd under a reflection of the chain. We will always assume 
$h \ge 0$. The eigenvalue equation for the single-particle states $| k \rangle$ then is
\eq{
 -\frac 1 2[\phi_k(n-1)+ \phi_k(n+1)]+h(n-1/2)\phi_k(n)=\omega_k\phi_k(n)
\label{eq:equXX},}
and its general solution is given by a linear combination of the Bessel functions
$J_{n-\kappa}(1/h)$ and $Y_{n-\kappa}(1/h)$. The argument of these functions 
defines the characteristic length $\lambda=1/h$, which will be of central importance in 
the following. For a finite system, $\kappa$ and $\omega_k=h(\kappa-1/2)$ follow from the 
boundary conditions.The resulting spectrum is shown on the left of Fig. 1 for $h=0.05$ 
and three values of $L$. One can see a linear region of equidistant levels with spacing 
$h$ in the center, while the level separation becomes larger at the upper and lower
end. Pictures of the corresponding eigenfunctions were first shown by Saitoh 
\cite{Saitoh73}. As $L$ increases, the regions with nonlinear dispersion are moved 
towards $\pm \infty$ and only the Wannier-Stark ladder with integer $\kappa=k$ remains.
The eigenfunctions then are $\phi_k(n)=J_{n-k}(1/h)$ and concentrated near 
site $k$ of the chain. In the ground state, the single-particle levels with $k \le 0$
are occupied. The resulting density profiles are shown on the right of Fig. 1 for
three values of the field. One can see that the transition from high to low density
takes place in a region of width $2\lambda$.
%
\begin{figure}[thb]
\center
\includegraphics[scale=.58]{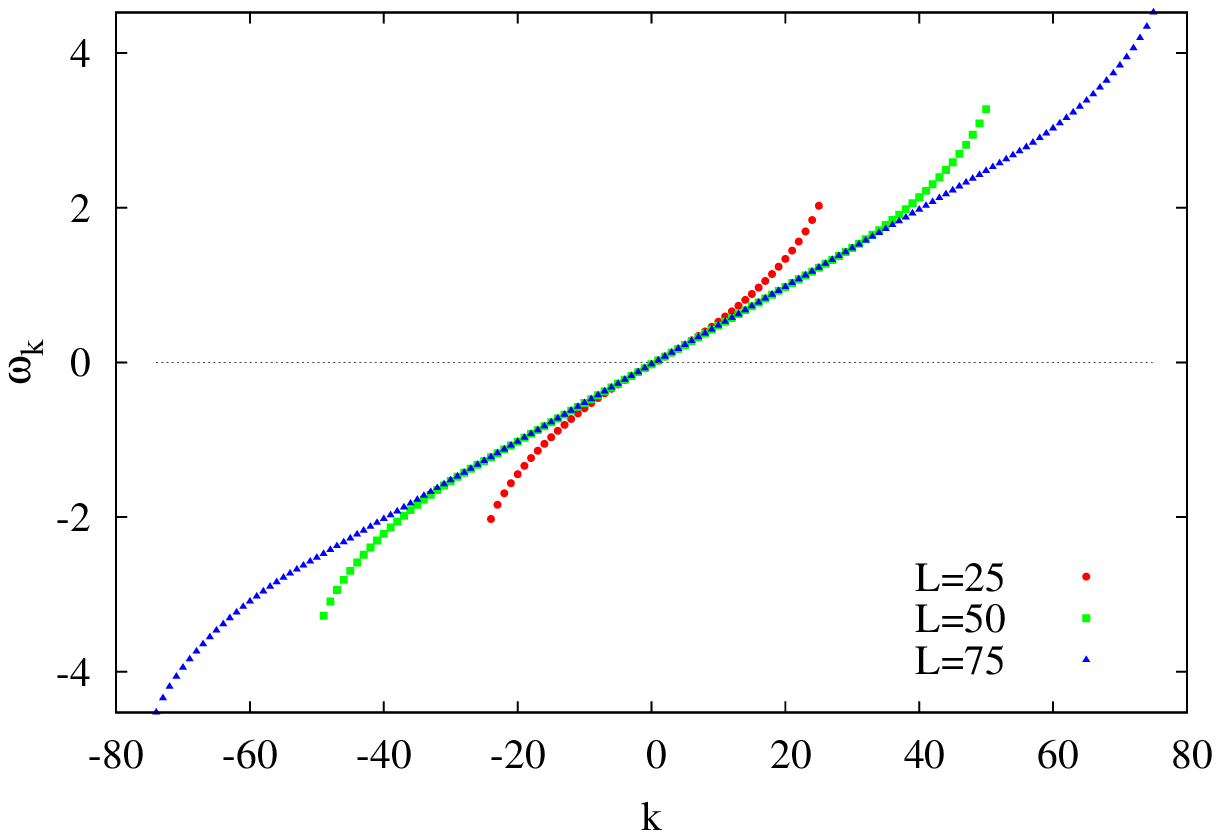}
\includegraphics[scale=.58]{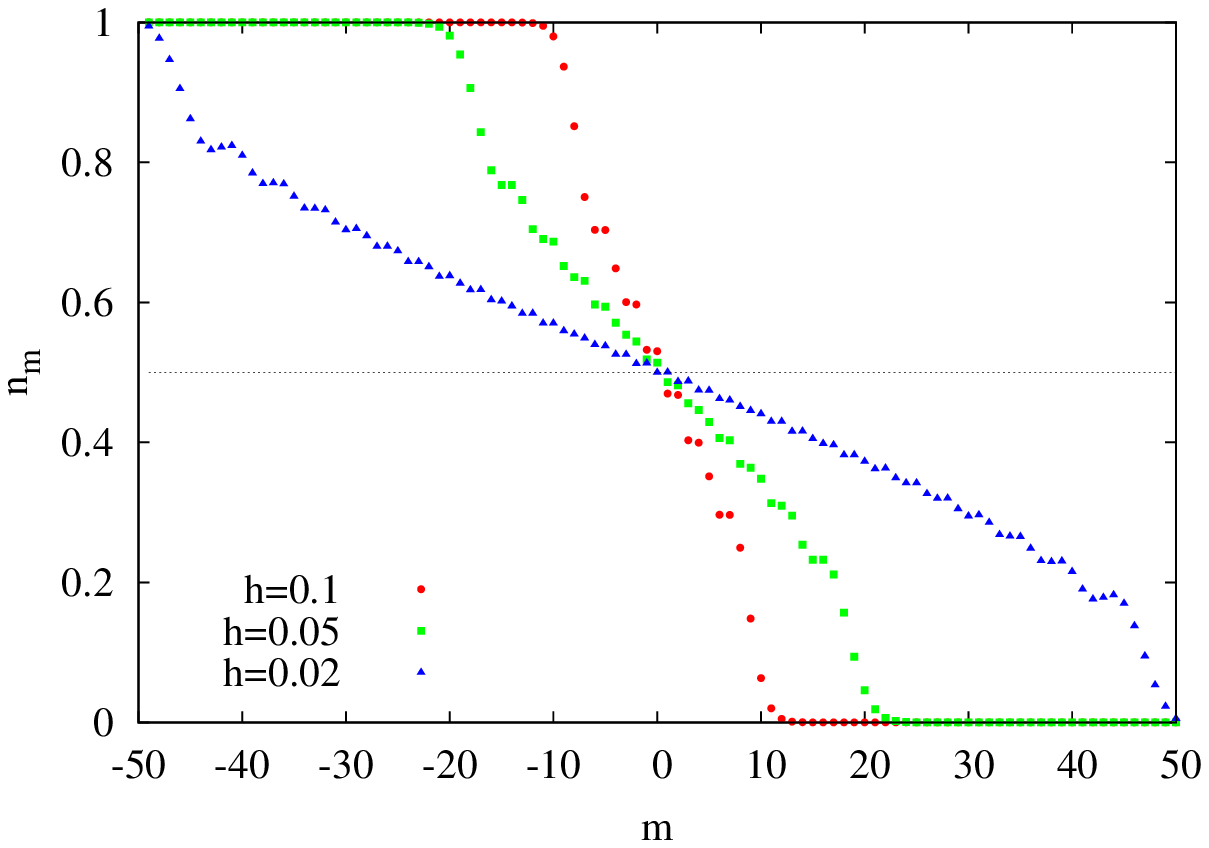}
\caption{Wannier-Stark problem. Left: Single-particle eigenvalues $\omega_k$ for $h=0.05$ 
and three values of L. Right: Density profiles in the ground state for $L=50$
and three values of the field.}
\label{fig:spectdensXX}
\end{figure}
%

\section{Entanglement in the XX chain}

In the following, we consider chains of $2L$ sites as in section 2 and study the 
entanglement between their left and right halves in the ground state. The corresponding 
entanglement entropy is $S= -\mathrm{Tr}\,(\rho \ln \rho)$ where $\rho$ denotes 
the reduced density matrix of one of the subsystems, e.g. the right half.
Both $\rho$ and $S$ follow \cite{Peschel03,Latorre03} from the correlation matrix 
\eq{
C_{mn}=\langle c^{\dag}_m c_{n} \rangle = \sum_{k=-L+1}^{L}
\phi_k(m) \phi_k(n) n_k, 
\label{eq:corrXX}}
where the $\phi_k(m)$ are the single-particle eigenfunctions appearing 
in (\ref{eq:equXX}) and $n_k$ the corresponding occupation numbers. 
In the ground state, these are one for $k \le 0$ and zero otherwise. Restricting 
the matrix to the sites of the subsystem, $1 \le m,n \le L$, and calculating 
its eigenvalues $\zeta_l$, one obtains $S$ as
\eq{
S =-\sum_l \zeta_l \ln \zeta_l -
\sum_l (1-\zeta_l) \ln (1-\zeta_l),
\label{eq:entropy}}
The calculation of the matrix and the diagonalization are done numerically.
\par
\begin{figure}[thb]
\center
\includegraphics[scale=.58]{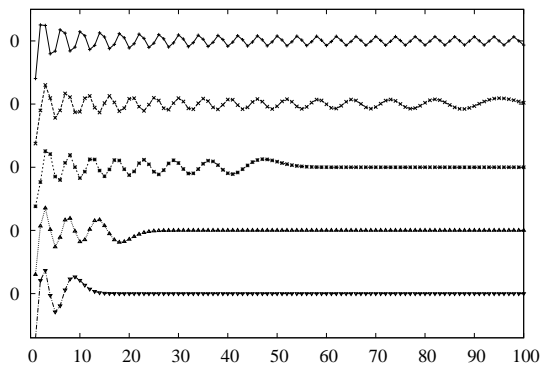}
\caption{Eigenvector of the correlation matrix for $L=100$ and five values of $h$.
From top to bottom: $h=0,\, 0.01,\, 0.02,\, 0.05,\, 0.1$. The center of the chain
is at the left end.}
\label{fig:evec_XX}
\end{figure}

   
We first show that the length scale $\lambda$ introduced by the gradient
appears directly in the single-particle eigenvectors of the correlation matrix.
In Fig. \ref{fig:evec_XX} we have plotted, for fixed $L$, the eigenvector 
corresponding to the $\zeta_l$ which is closest to $1/2$ and thus gives the 
largest contribution to $S$. In the homogeneous case, it decays slowly from the 
boundary between the subsystems but extends through the whole interior. If the 
gradient is large enough, however, it becomes confined to a region near the boundary
and the amplitude effectively vanishes at a distance $\lambda$. This is similar 
to the situation in a homogeneous non-critical system, for example a dimerized 
hopping model \cite{Greifswald}. In that case, the corresponding scale is the 
correlation length.
%
\par
\begin{figure}[thb]
\center
\includegraphics[scale=.7]{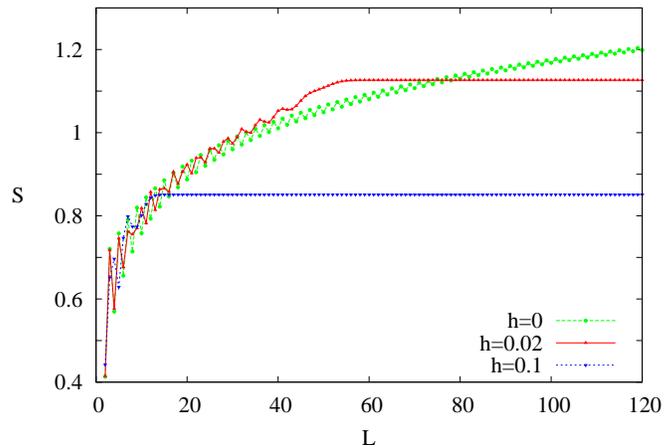}
\caption{Entanglement entropy for the XX chain with a linear field as a function
of the half-length $L$}
\label{fig:ent_XX}
\end{figure}

We now turn to the entanglement entropy. In Fig.  \ref{fig:ent_XX} it is shown 
as a function of $L$ for three values of $h$. For $h=0$ one has the
well-known logarithmic increase \cite{CC04}, but in a finite gradient 
$S$ bends over after an initial rise and saturates rapidly. 
The change in the behaviour takes place if $L \approx \lambda$.
This saturation is easy to understand since the parts of the system
outside the interface region are either full or empty and cannot
contribute to the entanglement.
%
\par 

\begin{figure}[thb]
\center
\includegraphics[scale=.7]{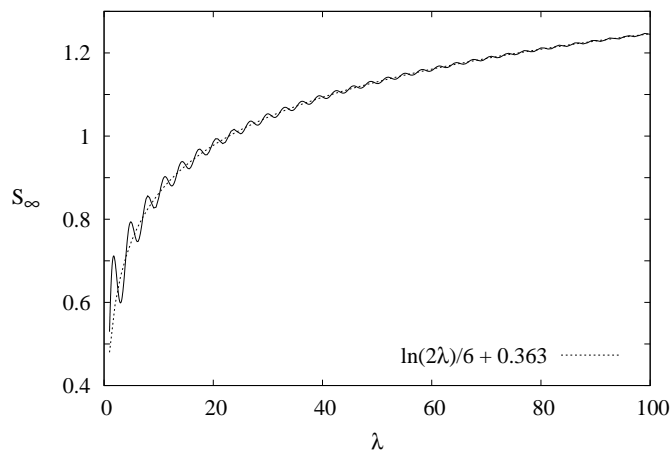}
\caption{Asymptotic value $S_{\infty}$ of the entanglement entropy as a function
of the length $\lambda$}
\label{fig:entasympt_XX}
\end{figure}

The saturation value, calculated numerically for sizes  $L=5 \lambda$,
is shown in Fig. \ref{fig:entasympt_XX} as a function of $\lambda$.
It varies essentially logarithmically, with additional decaying oscillations.
One can fit the data perfectly with the following form:
\eq{
S_{\infty}(\lambda)=\frac{1}{6}\ln (2\lambda)+\frac{k}{2}+ 
A \frac{\cos(2\lambda)}{\lambda}
\label{eq:entasympt_XX}}
where $A \approx 1/4$ and $k=0.726$ is the non-universal constant appearing
in the entanglement entropy of the \emph{homogeneous} XX chain of length 
$2L$, divided in the middle
\eq{
S_{\mathrm{hom}}=\frac{1}{6}\ln (4L/\pi)+\frac{k}{2} 
\label{eq:enthom_XX}}
The appearance of the constant $k$ in (\ref{eq:entasympt_XX}) is intriguing.
One could make the two formulae identical by introducing an effective length
$\lambda_{\mathrm{eff}}=\pi\lambda/2$ in (\ref{eq:entasympt_XX}), but this length
would not have the simple interpretation of an interface width. In section 5.1
it will be seen that there is also a close relation of $S_{\infty}(\lambda)$ to the 
time-dependent entropy after a certain quench.
%
\par
One should mention that in the case $L\to\infty$ the correlation matrix
can be written down analytically for any finite portion of the chain. 
As mentioned above, one then has to
deal only with the Wannier-Stark ladder states, where the eigenfunctions
are single Bessel functions of integer order. Thus 
\eq{
C_{mn}= \sum_{k=0}^{\infty} J_{k+m}(\lambda) J_{k+n}(\lambda)
\label{eq:corrXXlinf}}
Using a sum rule, (\ref{eq:corrXXlinf}) can be rewritten as
a simple product of Bessel functions. Dropping the arguments $\lambda$
it reads
\eq{
C_{mn} = \frac{\lambda}{2(m-n)} \left[J_{m-1} J_{n} - J_{m} J_{n-1} \right]
\label{eq:corrXXlinf2}}
In the limit $\lambda \to \infty$, this reduces to the well-known result
\eq{
C_{mn} = \frac{\sin(\pi(m-n)/2)}{\pi(m-n)} 
\label{eq:corrXXlinf3}}
of the homogeneous system. For general values of $\lambda$ it can easily be 
evaluated numerically. Since the Bessel functions become exponentially small 
for large order, taking $L \gg \lambda $ practically coincides with the result 
in the thermodynamical limit. Comparing with the numerical results using the
exact eigenfunctions of (\ref{eq:hamXX}) one also finds excellent agreement.
\par
\begin{figure}[thb]
\center
\includegraphics[scale=.7]{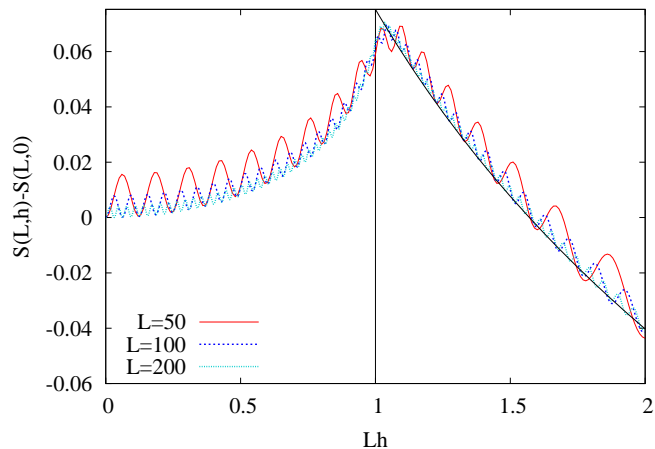}
\caption{Scaling behaviour of the difference $S(L,h)-S(L,0)$ 
as a function of $Lh$ for four system sizes. The full line in the right half
is the function (\ref{eq:scfun}).}
\label{fig:entscale_XX}
\end{figure}

We have also investigated the scaling limit, where the ratio of the 
length scales $L/\lambda=Lh$ is kept fixed while $L \to \infty$. We have 
calculated the entropy difference $\Delta S=S(L,h)-S(L,0)$ for several 
values of $L$. The results are shown in Fig. \ref{fig:entscale_XX}.
\par
%
%
In the case $Lh \gg 1$ we are in the Wannier-Stark limit discussed above, 
and for large $L$ the curves tend towards a scaling function $f(Lh)$
even for $Lh$ values only slightly above 1. This function is given
by the difference of the asymptotic $\lambda \to \infty$ form of
(\ref{eq:entasympt_XX}) and the formula (\ref{eq:enthom_XX}) for the entropy
of a half-chain. Taking the difference one finds
\eq{f(x)=\frac{1}{6}\ln\frac{\pi}{2x}\label{eq:scfun}}
with $x=Lh$. The maximum is  
$\Delta S_{\mathrm{max}} = f(1) = \frac{1}{6}\ln\frac{\pi}{2}=0.0753$ in agreement
with the value in the figure, and thus relatively small. It mirrors the slight 
bulge in the curve for $S(L)$ seen in Fig. \ref{fig:ent_XX} before the asymptotic
value is reached.
\par
For $Lh \le 1$ the interface region fills the whole system, but the curves still 
show a nice scaling behaviour and $\Delta S$ approaches zero quadratically as 
$Lh \to 0$.

\section{Transverse Ising chain}

We now consider the inhomogeneous quantum Ising chain with Hamiltonian
\eq{
H=-\sum_{n=-L+1}^{L-1} J_n \sigma_n^z\sigma_{n+1}^z-h\sum_{n=-L+1}^{L} \sigma_n^x
\label{eq:hamTI}}
where $\sigma_n^x$ and $\sigma_n^z$ are the components of a Pauli spin operator associated 
with site $n$,  $J_n$ is the nearest-neighbour exchange interaction and $h$ the 
transverse field. The couplings are assumed to vary as 
\eq{
J_n=J[1+gn]\;.
\label{eq:coupTI}}
The homogeneous chain with $g=0$ has a quantum critical point at $J=h$ in the 
thermodynamic limit, $L \to \infty$. Thus setting $h=J$, the inhomogeneous
system with $g > 0$ is undercritical in the left half and overcritical, i.e.
in the ordered phase, on the right.
The length characterizing the transition region can be obtained from a scaling
argument given in the Appendix \cite{Platini07}. This leads to 
$\lambda(g)=ag^{-\omega}$, for small $g$. Here $\omega=\nu/(1+\nu)$ where $\nu$ 
is the correlation length exponent. It enters because the perturbation drives 
the system away from criticality. With $\nu=1$ for the TI model and choosing the 
constant $a$ equal to one, $\lambda(g)=g^{-1/2}$. The result for the XX 
chain can also be obtained in this way by using $\nu=\infty$ for the marginal 
perturbation one has there.
\par
\begin{figure}
\begin{center}
\includegraphics[scale=.6]{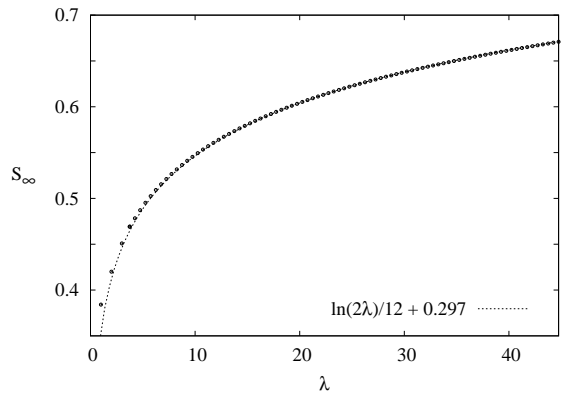}
\end{center}
\caption{Limiting value of the entanglement entropy in a TI chain 
as a function of $\lambda(g)=g^{-1/2}$. For comparison we have also plotted
$1/12 \ln x + k_1$ with $k_1=0.297$.}
\label{fig:entasympt_TI}
\end{figure}

The entanglement entropy between the two halves of the chain in its ground state is 
calculated again from the correlation functions by writing (\ref{eq:hamTI}) in terms 
of fermions. Here one can work either in terms of Majorana operators \cite{Latorre03}
or of Fermi operators \cite{IgloiLin08}. For $S$ as a 
function of $L$ one finds the same overall behaviour as in the XX model, i.e. it 
rises logarithmically for small $L$ and saturates then. The main difference is the 
absence of additional oscillations. This also holds for the asymptotic values which 
are plotted in Fig. \ref{fig:entasympt_TI} as a function of $g^{-1/2}$.
%
%
The function $S_\infty$ can be fitted by a simple logarithm
\eq{
S_{\infty}(\lambda)=\frac{1}{12}\ln (2\lambda)+k_1 
\label{eq:entasympt_TI}}
The factor $1/12$ instead of $1/6$ corresponds to the central charge $c=1/2$ of 
the TI model. The constant is given by $k_1=0.297$. In contrast to the XX case, 
there seems to be no relation to the constant 0.239 appearing in the entropy
of the homogeneous model.
\par
\begin{figure}
\begin{center}
\includegraphics[scale=.65]{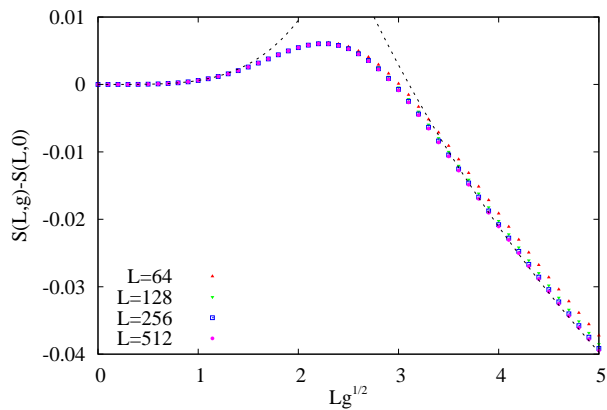}
\end{center}
\caption{The entropy difference S(L,g)-S(L,0) as a function of $x=Lg^{1/2}$
for several values of $L$. The lines show the limiting behaviour, see text.}
\label{fig:entscale_TI}
\end{figure}

With this information, we can again analyze the finite-size behavior of the entropy,
which is expected to be 
\eq{
S(L,g)-S(L,0)= f(L/\lambda).
\label{eq:scaling}}
Indeed, for large $L$ the entropy-difference approaches a universal function,
as shown in Fig.\ref{fig:entscale_TI}.
%
%
%
For small argument the scaling function behaves as $S(x) \sim x^4 \sim g^2$, which 
follows from the fact that the entropy must be an even function of $g$.
For large arguments the asymptotic behaviour is
$f(x) \simeq - 1/12 \ln x+\mathrm{const}$
which is in agreement with the result in Eq.(\ref{eq:entasympt_TI}). For intermediate 
values of $x$, the scaling function has a maximum at $x_{\mathrm{max}} \approx 4.6$
but, in contrast to the XX case in Fig. \ref{fig:entscale_XX}, is non-singular for 
all finite values of $x$.
\par
Instead of having the gradient in the couplings, one can also put it into the transverse
field. In this case, due to duality, the ordered and disordered sides of the chain
are reversed, but the size of the interface $\lambda(g)$ is expected from scaling 
theory to vary in the same way as before. Calculating the entanglement entropy, it is
seen that $S$ has again the form (\ref{eq:entasympt_TI}) and also the scaling 
function in (\ref{eq:scaling}) approaches for large $L$ a universal function with
the same type of limiting behaviour.
\par
\begin{figure}
\center
\psfrag{l/2L}[][][.6]{$\ell$/2L}
\psfrag{S(l)}[][][.6]{S($\ell$)}
\includegraphics[scale=.6]{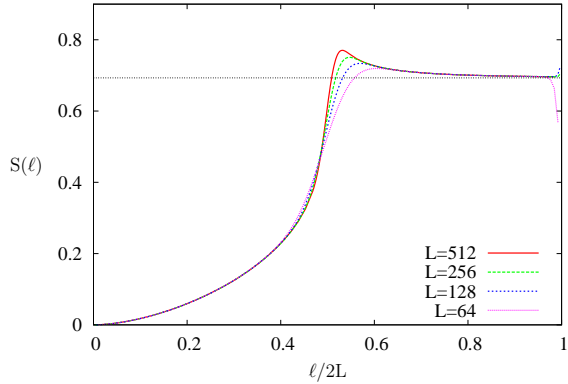}
\caption{Entanglement entropy for TI chains divided asymetrically in two
segments of length $\ell$ and $2L-\ell$ for several system sizes and with $g=1/L$.}
\label{fig:ent_blocksize_TI}
\end{figure}

The three regimes of the system in a gradient, namely disordered (paramagnetic), 
interfacial and ordered (ferromagnetic) can also be probed by dividing the chain not in
the middle but in two parts of length $\ell$ and $2L-\ell$ and calculating the 
corresponding entanglement entropy. In Fig. \ref{fig:ent_blocksize_TI}
we present results for the case, where $g=1/L$ which means that the coupling
$J_n$ vanishes at the left boundary and equals $2$ at the right one. 
\par
%
%
\par
Both in the paramagnetic and in the ferromagnetic regime, the entropies
approach a master curve as $L$ is increased. In the interface regime, the entropy
has a maximum which is of the same order as the value for $\ell=L$. The
shape of the curves is reminiscent of those one finds in a \emph{homogeneous} 
system of different finite sizes, if one varies the coupling constant \cite{CC04}. 
In fact, if one uses the coupling $J_{\mathrm{hom}}(\ell)=J(1-\ell/L)$ in the 
homogeneous system and calculates $S$ for a chain divided in the middle, the 
resulting curves are very close to those shown in 
the figure, even in the interface region, provided the size of the homogeneous system 
is close to the size of that region. This equivalence can be understood as follows.
The length scale in the gradient system is $\lambda$, thus the correlation between 
two points having a distance larger that $\lambda$ vanishes. Consequently, the
contribution to the entropy at a given separation point $x$ is coming from the
sites in the range $[x-\lambda,x+\lambda]$. In this correlated domain, the
couplings are varying only weakly and can be replaced by their average, which is
just $J_{\mathrm{hom}}(\ell)$. Therefore the entropy is that of the homogeneous system
with size  $\lambda$ and separation point in the middle.


\section{Time evolution after a quench}

The equilibrium results show that the main effect of the gradient is the 
introduction of a length scale, in terms of which the entropy still shows a 
logarithmic scaling. We now ask what happens if the gradient is suddenly 
switched off and a non-trivial time evolution of the state sets in.
\par
\subsection{XX chain}

In the XX case, the fermionic operators evolve after the quench according 
to \cite{EP07}
\eq{c_j(t)=\sum_m U_{jm}(t) c_m \, , \quad 
U_{jm}(t)= \sum_q \psi_q(j)\psi_q(m)\ee^{it \cos q}.
\label{eq:unitXX}}
where the sum is over the allowed momenta $q=\pi k/(2L+1), k=1,2,\dots 2L$ for
the homogeneous open chain, $\psi_q(j)= L^{-1/2} \sin(q(j+L))$ are the 
single-particle eigenfunctions and 
$\omega_q=-\cos q$ the corresponding eigenvalues. Therefore the correlation
matrix ${\bf C}(t)$ at time $t$ is obtained by multiplying ${\bf C}(0)$ from 
the left and right by the matrices ${\bf U^{\dag}}(t)$ and ${\bf U}(t)$, 
respectively. The entropy is then calculated from ${\bf C}(t)$ as before.
\par
\begin{figure}[thb]
\center
\includegraphics[scale=.7]{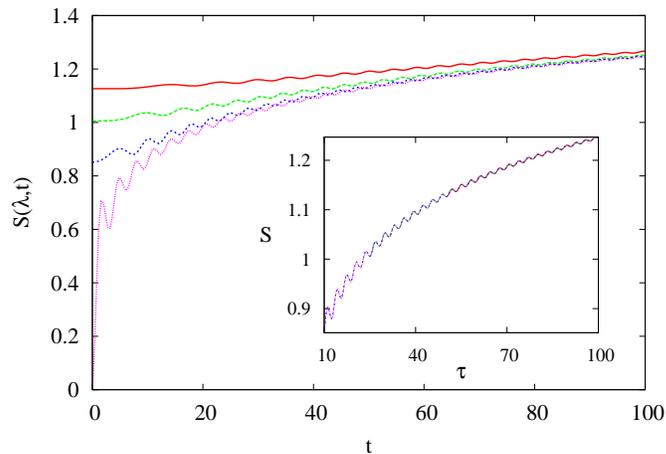}
\caption{Time evolution of the entanglement entropy in an XX chain with $L=150$
after switching off the gradient. The initial values were, from top to
bottom $\lambda = 50,20,10,0$. The inset shows $S$ as a function of the variable
$\tau=\sqrt{t^2+\lambda^{2}}$}
\label{fig:entevol_XX}
\end{figure}

In Fig. \ref{fig:entevol_XX} we show the resulting time evolution of $S$ 
for various values of the initial gradient. For $\lambda=0$ 
one is starting from a perfectly sharp domain wall. This situation has
already been studied with regard to the evolution of the density in
\cite{ARRS99,HRS04} and with respect to the particle-number fluctuations
in \cite{Antal08}. The entanglement entropy was obtained in a DMRG
calculation in \cite{Gobert05}, but not investigated further. The
curve looks very much like the one in Fig. \ref{fig:entasympt_XX},
on which we will comment presently. For larger $\lambda$ values,
the initial entropy is higher, but the time evolution is
also slower, and asymptotically all the curves seem to converge
to the one with $\lambda=0$. Moreover, introducing the new variable
$\tau=\sqrt{t^2+\lambda^{2}}$ one finds an \emph{exact} collapse of
the data, as shown in the inset.
%
\par
This result can be derived analytically if one considers the limit 
$L\to\infty$. In this case, one can work with a ring instead of an open
chain and the quantities $U_{jm}$ become Bessel functions. The equation
for ${\bf C}(t)$ then reads explicitly  
\eq{C_{jl}(t)=i^{l-j} \sum_{m,n} i^{m-n} J_{j-m}(t) J_{l-n}(t) C_{mn}(0)
 \, ,\label{eq:corrtXX}}
Furthermore the matrix ${\bf C}(0)$ is given by (\ref{eq:corrXXlinf}) in
this limit. Therefore one has sums of Bessel functions with two different 
arguments, $t$ and $\lambda$. Using their integral representations, the
infinite sums over $m$ and $n$ can be carried out and one obtains
\eq{C_{jl}(t)=\sum_{k\ge 0} F^{*}_{j+k}(\lambda,t) \, F_{l+k}(\lambda,t)
\label{eq:corrtXX2}}
where
\eq{F_{n}(\lambda,t)=\int_{-\pi}^{\pi}\frac{\dd q}{2\pi}
\ee^{it\cos q - i\lambda\sin q + iqn}.\label{eq:genbes}}
In this integral one can now rewrite the variables as 
$\lambda =\tau \cos \varphi$ and $t=\tau \sin \varphi$. The addition 
theorem for the trigonometric functions then yields
\eq{F_{n}(\tau,\varphi)= \int_{-\pi}^{\pi}\frac{\dd q}{2\pi}
\ee^{-i\tau\sin (q-\varphi) + iqn} = \ee^{i \varphi n} J_n(\tau)
\label{eq:genbes2}.}
with $\varphi=\arctan (t/\lambda)$.
\par
Thus, up to phase factors the quantities $F_n$ are Bessel functions with
argument $\tau$, a fact which was not realized in \cite{CaseLau73}. 
At the level of the correlation matrix, the phase factors correspond to
a simple unitary transformation $\bf{C} \to \bf{U^\dag}\bf{C}\bf{U}$ 
and do not affect the entanglement entropy for which one finds the relation
\eq{S(\lambda,t)=S(0,\sqrt{t^2+\lambda^2})\label{eq:isoent}}
The entropy thus depends only on the variable $\tau$ which is simply
the distance from the domain-wall initial state in a space-time coordinate
system where in the space direction we move to the equilibrium system with
interface length $\lambda$ and then we further evolve this state in time.
The lines of constant entropy are therefore circles in this quarter-plane.
This explains in particular the result mentioned above that $S(\lambda,0)=
S(0,t)$ for $t \to \lambda$.  Alternatively, for arbitrary time $t$ one could
think of the evolving state as being effectively the ground state of
a gradient problem with interface length $\lambda_{\mathrm{eff}}(t)=
\sqrt{\lambda^{2}+t^2}$. Hence, one has a front propagating with a
time-dependent speed $v(t)=\dd \lambda_{\mathrm{eff}} / \dd t$.
\par
Before closing this section we note that in finite systems one finds 
additional features at times $\tau=2L,4L,\dots$, which are larger than those 
in Fig. \ref{fig:entevol_XX}. Then the entropy shows a step-like increase which
can be attributed to the propagating fronts which return to the center
after being reflected at the open ends \cite{EKPP08}.
\par
%
\subsection{TI chain}
\par
\begin{figure}[thb]
\center
\includegraphics[scale=.7]{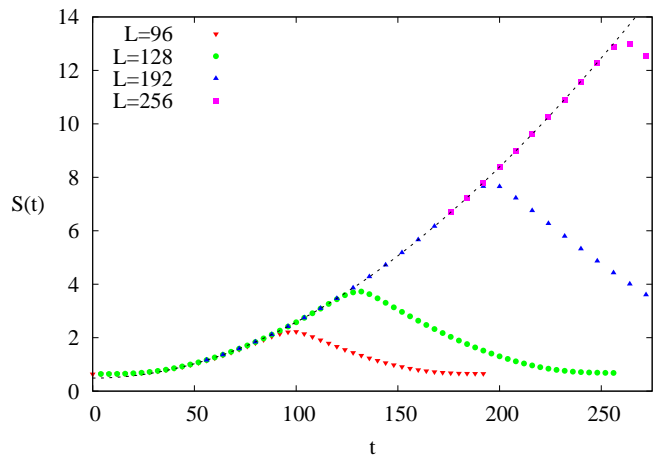}
\caption{Time evolution of the entanglement entropy in the TI chain
after switching off a gradient $g=1/1024$, for several values of $L$.
The broken line is the fit using (\ref{eq:entevol_TI}).}
\label{fig:entevol_TI}
\end{figure}

In the TI chain, the calculation of the correlations using Majorana operators 
proceeds basically in the same way, but the time-dependent factors now contain
the excitation energies $\omega_q=2\sin(q/2)$ of the critical homogeneous 
system with $h=J=1$. This gives a maximum velocity $v=1$ for the excitations.
In Fig. \ref{fig:entevol_TI} the resulting entropy is shown for $g=1/1024$ corresponding
to $\lambda =32$, and four different lengths which were all much larger than $\lambda$.
One can see that $S$ increases up to $t \sim L$ and then drops again. The increase
is different from the linear law one finds in homogeneous systems and can be 
fitted by a form 
\eq{
S(t,g)= a(g) t^2 - b(g) t^3 + c(g)
\label{eq:entevol_TI}}
where the cubic term represents a slight modification of the quadratic law.
We have written $a(g),\, b(g)$ and $c(g)$, because for other $g$
one finds the same time behaviour but with different coefficients.
In particular $a(g)$ is approximately linear in $g$, i.e. $a(g)= a g$
where $a \simeq 0.23$.

%
%
%
%
The behaviour found above can be understood using the picture developed in 
\cite{CC05} that at the quench pairs of quasiparticles are emitted which 
establish the entanglement between the parts of the system at later times. The
number of these pairs must depend on the ``distance'' of the initial state
from the ground state of the final system. This has to be connected with the
difference between the initial coupling constants and the critical value $J=1$.
The precise expression can be found by considering a quench from a homogeneous
initial system with $J \ne 1$ and determining the coefficient $\alpha$ in 
the linear law $S(t)-S(0)=\alpha t$. This can be done numerically for our finite 
geometry, or by using the formula (2) in \cite{Fagotti08} for a segment in an 
infinite chain and dividing the result by 2. The latter approach gives 
$\alpha$ as an integral over momenta
\eq{
\alpha(J) =\int_{-\pi}^{\pi} \frac {dq}{2\pi} v_q H(y_q)
\label{eq:alpha1}}
where $v_q=\cos(q/2)$ is the velocity of the final quasiparticles,
\eq{
y_q = \frac {(J+1) \sin(q/2)}{\sqrt{(J-1)^2+4J \sin^2(q/2)}}
\label{eq:ypsilon}}
and
\eq{
H(y) = -\left[\frac {1+y}{2} \ln(\frac {1+y}{2}) 
   + \frac {1-y}{2} \ln(\frac {1-y}{2})\right].
\label{eq:function_H}}
Using partial integrations, this can be evaluated in closed form and gives
\eq{
\alpha(J)= \frac {|J-1|}{\pi\sqrt{J}}\left[ \frac {\pi}{2}- 
 \arctan(\frac {|J-1|}{2\sqrt{J}}) \right]- \frac {(J-1)^2}{2\pi J}
\ln \left| \frac {J+1}{J-1} \right|.
\label{eq:alpha2}}
For small $|J - 1|$ this varies linearly, $\alpha \simeq |J - 1|/2$, while for 
$J \to  \infty$ (or for $J \to 0$) it approaches the saturation value $1/\pi$. 
The last term in (\ref{eq:alpha2}) is non-analytic at $J=1$, but a good 
approximation in the range $|J - 1| \lesssim 0.2$ is
\eq{
\alpha(J) = 0.475 |J - 1|-0.55(J - 1)^2
\label{eq:alpha}}
which we will use for simplicity. A similar result follows from the direct 
calculation in the finite geometry.
\par
The quantity $\alpha/2$ can now be used as the density of emitted pairs in the 
phenomenological formula of Calabrese and Cardy, which gives $S(t)$, in a
continuum approximation, as
\eq{
S(t)- S(0) = \frac {1}{2} \int_{-t}^{t} \dd x \, \alpha(J) = \alpha(J) t
\label{eq:entpheno1}}
where the integral counts all the pairs which end up in different halves of the
system up to time $t$. In an \emph{inhomogeneous} system, the obvious generalization 
of this formula is
\eq{
S(t)-S(0)= \frac {1}{2} \int_{-t}^{t} \dd x \, \alpha(J(x))
\label{eq:entpheno2}}
If the initial couplings vary along the chain as $1 + gn^{\theta}$, this gives
\eq{
S(t)-S(0)=  \frac {0.475}{\theta +1} g \, t^{\theta +1} - 
\frac {0.55}{2\theta +1} g^2 \, t^{2\theta +1}
\label{eq:entpheno3}}
In the case $\theta=1$, this has exactly the form (\ref{eq:entevol_TI}) and moreover
the coefficient of the $g t^2$ term equals 0.24 which is very close to the value 0.23
found by fitting. The agreement is also good for the cubic term where $b(g)/g^2$ equals
0.18 in (\ref{eq:entpheno3}) and 0.12 in the fit. Hence this formula describes the
increase of $S$ very well. In order to check it further, we have also studied the
case $\theta=2$, where the couplings increase quadratically from the center. Then 
$S(t)$ looks very similar, but a closer analysis shows that it varies basically as 
$g t^3$, which is again the prediction of (\ref{eq:entpheno3}). Also the coefficient
has the correct value. Nevertheless one should
point out that there is a region of small times where the numerical $S(t)$ is rather
flat and not so well described by the formulae. This seems to hold up to $t \sim \lambda$
and would mean that inside the interface region the picture has to be modified.
Another remark concerns the form of Eq. (\ref{eq:entpheno2}). To leading order, it is 
an integral over $|J - 1|$, which can also be interpreted as the (local) energy
gap in the initial state. Such an expression was also used in a recent field theoretical 
treatment of inhomogeneous quenches, see Eq. (76) in \cite{Sotiriadis08}. In our approach
we also obtain the exact prefactor on the lattice. However, the integral (\ref{eq:alpha1})
from which it follows, does not permit to read off the result directly.
\par
Finally, let us comment on the decrease of $S(t)$ beyond $t=L$. This is a feature of the
finite geometry which one also finds in homogeneous quenches.
There a linear increase of $S$ is followed by an almost linear decrease and a zig-zag
variation of $S(t)$ results
for larger times. It can also be understood in the quasiparticle picture. Due to the 
open ends, the quasiparticles moving outwards are reflected at the ends and follow 
their inward moving partners with a certain delay. As soon as they cross the middle, 
their contribution to the entanglement between left and right vanishes. 
This effect sets in when the quasiparticles from the ends arrive at the center,
because their partners follow immediately.

\section{Summary}

We have studied particular inhomogeneous systems where a power-law variation of some
parameter introduces an interface with a certain width $\lambda$. Beyond that region the
ground state approaches a product form. This suggests that instead of the real length
of the system only $\lambda$ should enter the entanglement properties. In fact, we
found that both for the XX model and the TI model the entropy in large systems is given 
by conformal expressions where $\lambda$ appears in the logarithms. This result is also 
plausible because one knows that the entanglement is connected with the interface
between the two subsystems one considers. One finds the same $\ln \lambda$ behaviour
in the $q$-symmetric XXZ Heisenberg chain, where the interface is produced by boundary
fields \cite{SandowSchutz94} and the reduced density matrix is known explicitly
\cite{KaulkePeschel98}. It is also somewhat similar to the situation
for non-critical states, where the correlation length appears in the formulae. However, 
in our case, there is no translational invariance. For finite lengths $L$ we have
also shown that the entanglement entropy has a scaling form in the variable $L/\lambda$.
\par
The time evolution after a removal of the gradient turned out to be particularly 
interesting. In the XX case, it is logarithmic as found usually in \emph{local}
quenches \cite{EP07,CC07,EKPP08}. Moreover, it displays a particular
space-time symmetry relating static and dynamic entanglement. Formally, this results
because both the single-particle states in the Wannier-Stark problem and the time
evolution in the final homogeneous system are given by Bessel functions. Thus if one
can treat, for example, the quench from a sharp interface, one has found the solution 
for all gradients and all times. However, one cannot apply the CFT approach of  
\cite{CalaHaLeDous08} since the walls there are of a different nature and lead
to a linear behaviour of $S(t)$.
\par
In the TI model, the time evolution after the quench turned out to be quadratic,
a result not encountered before. We were able to explain this in the simple 
quasiparticle picture of \cite{CC05} and could even give
the numerical constants. We mentioned that more general power-law variations
of the couplings lead to analogous results which can be understood in the same way.
In a sense, this explanation works better than one might expect, because the
assumption that quasiparticles are only emitted from nearby sites is not well
founded near a critical point. Thus a more direct derivation following the 
lines of \cite{Sotiriadis08} would certainly be useful and interesting.

\section*{Acknowledgement}

F.I. is indebted to the Freie Universit\"at Berlin for hospitality during the 
starting period of this research. His work has been supported by the 
Hungarian National Research Fund under grant No. OTKA
TO48721, K62588 and K75324.

\section*{Appendix}

Although our main concern in this study were linearly varying parameters,
similar results are obtained for other power laws. Consider a variation 
of the couplings in the transverse Ising model of the form
\eq{
J_n=J[1+g|n|^{\theta}]\;.
\label{eq:gencoupTI}}
This leads to a length scale $\lambda$ which can be estimated as follows.
The typical deviation of the couplings from the critical value is 
$\Delta (\lambda) \sim J g\lambda^{\theta}$, which leads to a length scale
$\xi \sim \Delta (\lambda)^{-\nu}$. Since in the problem there is only one 
length scale, the width of the interface, we have $\xi \sim \lambda$ from 
which one obtains the self-consistency equation
\eq{
\lambda \sim \left[g\lambda^{\theta}\right]^{-\nu}\;,
\label{eq:lambda1}}
with the solution:
\eq{
\lambda= a\;g^{-\nu/(\theta \nu+1)}\;.
\label{eq:lambda2}}
Using $\nu=1$ the exponent is $\omega=1/(\theta+1)$ which gives $1/2$
for the linear variation considered in the main text.
\par
\begin{figure}
\begin{center}
\includegraphics[scale=0.6]{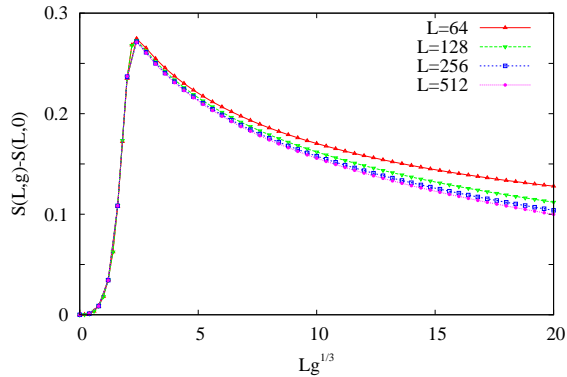}
\end{center}
\caption{Entropy difference $S(L,g)-S(L,0)$ as a function of $Lg^{1/3}=L/\lambda$ for a
quadratic variation of the couplings in the TI model.}
\label{fig:entscale_quad_TI}
\end{figure}

As an example, let us consider a quadratic variation, $\theta=2$.
Physically, this means that for $J=h$ and $g \ge 0$ the system is critical
in the center and ordered more and more as one moves towards the ends. Thus
one has a kind of sandwich structure with the thickness of the central part
varying as $\lambda \sim g^{-1/3}$. Calculating the entropy, one finds again
that it saturates for large $L$ and the asymptotic value varies as in (\ref
{eq:entasympt_TI}). The constant now has the value $k_2=0.548$ if one sets
$a=1$ in (\ref{eq:lambda2}). The difference $S(L,g)-S(L,0)$
is shown in Fig. \ref{fig:entscale_quad_TI}.
%
%
%
The scaling function resembles that of the XX model in that it also has a cusp
separating two different regimes. For small arguments it varies as 
$x^3$, i.e. it is proportional to $g$. This behaviour
is possible here since $g<0$ and $g>0$ are not equivalent.
For large arguments it is given by $-1/12 \ln x + \mathrm{const}$.
\par
The case $\theta=1$ corresponding to a linear ``trapping potential'' for the
central part gives very similar results. The length is now $\lambda=g^{1/2}$ as
in section 4, but the scaling function looks qualitatively as in Fig.
\ref{fig:entscale_quad_TI}. However, for small arguments the behaviour
is quadratic.

\section*{References}



\begin{thebibliography}{10}
\expandafter\ifx\csname url\endcsname\relax
  \def\url#1{{\tt #1}}\fi
\expandafter\ifx\csname urlprefix\endcsname\relax\def\urlprefix{URL }\fi
\providecommand{\eprint}[2][]{\url{#2}}

\bibitem{Amicoetal08}
Amico L, Fazio R, Osterloh A and Vedral V, 2008 {\em Rev. Mod. Phys.\/} {\bf 80}
  517

\bibitem{CC04}
Calabrese P and Cardy J~L, 2004 {\em J. Stat. Mech.\/} P06002

\bibitem{Peschel05}
Peschel I, 2005 {\em J. Phys. A: Math. Gen.\/} {\bf 38} 4327

\bibitem{Levine08}
Levine G~C and Miller D~J, 2008 {\em Phys. Rev. B\/} {\bf 77} 205119

\bibitem{Levine04}
Levine G~C, 2004 {\em Phys. Rev. Lett.\/} {\bf 93} 266402

\bibitem{ZhaoPeWa06}
Zhao J, Peschel I and Wang X, 2006 {\em Phys. Rev. B\/} {\bf 73} 024417

\bibitem{RefaelMoore04}
Refael G and Moore J~E, 2004 {\em Phys. Rev. Lett.\/} {\bf 93} 260602

\bibitem{Laflo05}
Laflorencie N, 2005 {\em Phys. Rev. B\/} {\bf 72} 140408 (R)

\bibitem{IgloiLin08}
Igl\'oi F and Lin {\relax Y-Ch}, 2008 {\em J. Stat. Mech.\/} P06004

\bibitem{Wannier60}
Wannier G~H, 1960 {\em Phys. Rev.\/} {\bf 117} 432

\bibitem{Smith71}
Smith E~R, 1971 {\em Physica\/} {\bf 53} 289

\bibitem{Saitoh73}
Saitoh M, 1973 {\em J. Phys. C: Solid State Physics\/} {\bf 6} 3255

\bibitem{CaseLau73}
Case K~M and Lau C~W, 1973 {\em J. Math. Phys.\/} {\bf 14} 720

\bibitem{SteyGusman73}
Stey G~C and Gusman G, 1973 {\em J. Phys. C: Solid State Physics\/} {\bf 6} 650

\bibitem{Wilkinson96}
Wilkinson S~R, Bharucha C~F, Madison K~W, Niu Q and Raizen M~G, 1996 {\em Phys.
  Rev. Lett.\/} {\bf 76} 4512

\bibitem{Niu96}
Niu Q, Zhao {\relax X-G}, Georgakis G~A and Raizen M~G, 1996 {\em Phys. Rev.
  Lett.\/} {\bf 76} 4504

\bibitem{Platini07}
Platini T, Karevski D and Turban L, 2007 {\em J. Phys. A: Math. Theor.\/} {\bf
  40} 1467

\bibitem{CC07}
Calabrese P and Cardy J~L, 2007 {\em J. Stat. Mech.\/} P10004

\bibitem{Feuer52}
Feuer P, 1952 {\em Phys. Rev.\/} {\bf 88} 92

\bibitem{PRS94}
Peschel I, Rittenberg V and Schultze U, 1994 {\em Nucl. Phys. B\/} {\bf 430
  [FS]} 633

\bibitem{Peschel03}
Peschel I, 2003 {\em J. Phys. A: Math. Gen.\/} {\bf 36} L205

\bibitem{Latorre03}
Latorre J~I, Rico E and Vidal G, 2004 {\em Quantum Inf. Comput.\/} {\bf 4} 48

\bibitem{Greifswald}
Peschel I and Eisler V, in: Computational Many-Particle Physics, 
Fehske H, Schneider R and Weisse A, eds 2008 {\em Lecture Notes in Physics\/}
vol 739 (Springer Berlin) pp 581-596

\bibitem{EP07}
Eisler V and Peschel I, 2007 {\em J. Stat. Mech.\/} P06005

\bibitem{ARRS99}
Antal T, R\'acz Z, R\'akos A and Sch\"utz G~M, 1999 {\em Phys. Rev. E\/} {\bf
  59} 4912

\bibitem{HRS04}
Hunyadi V, R\'acz Z and Sasv\'ari L, 2004 {\em Phys. Rev. E\/} {\bf 69} 066103

\bibitem{Antal08}
Antal T, Krapivsky P~L and R\'akos A, 2008 Preprint arXiv:0808.3514

\bibitem{Gobert05}
Gobert D, Kollath C, Schollw\"ock U and Sch\"utz G, 2005 {\em Phys. Rev. E\/}
  {\bf 71} 036102

\bibitem{EKPP08}
Eisler V, Karevski D, Platini T and Peschel I, 2008 {\em J. Stat. Mech.\/} P01023

\bibitem{CC05}
Calabrese P and Cardy J~L, 2005 {\em J. Stat. Mech.\/} P04010

\bibitem{Fagotti08}
Fagotti M and Calabrese P, 2008 {\em Phys. Rev. A\/} {\bf 78} 010306(R)

\bibitem{Sotiriadis08}
Sotiriadis S and Cardy J, 2008 Preprint arXiv:0808.0116

\bibitem{SandowSchutz94} Sandow S and Sch\"utz G, 1994  {\em Europhys. Lett.} {\bf 26} 7

\bibitem{KaulkePeschel98} Kaulke M and Peschel I, 1998 {\em Eur. Phys. J. B\/} {\bf 5} 727

\bibitem{CalaHaLeDous08}
Calabrese P, Hagendorf C and Le Doussal P, 2008 {\em J. Stat. Mech.\/} P07013

\end{thebibliography}

\providecommand{\newblock}{}

\end{document}